\documentclass{article}
\usepackage{spconf,amsmath,graphicx}
\makeatletter
\def\ninept{%
  \def\baselinestretch{1.05}
  \renewcommand\normalsize{%
    \@setfontsize\normalsize{9.7pt}{9.7pt}%
  }%
  \normalsize
}
\makeatother

\usepackage{amsmath,amssymb,amsfonts}
\usepackage{algorithmic}
\usepackage[ruled,vlined]{algorithm2e}
\usepackage{graphicx}
\usepackage{textcomp}
\usepackage{tikz}
\usetikzlibrary{calc,arrows.meta} 
\usepackage{enumitem}

\usepackage{comment}
\usepackage{xcolor}
\usepackage{subcaption} 
\usepackage[table]{xcolor}
\def\BibTeX{{\rm B\kern-.05em{\sc i\kern-.025em b}\kern-.08em
    T\kern-.1667em\lower.7ex\hbox{E}\kern-.125emX}}

\SetKwComment{Comment}{$\triangleright$\ }{}

\usepackage{booktabs}
\usepackage{multirow}
\usepackage{siunitx} 
\usepackage{pifont} 
\usepackage{fontawesome5} 
\usepackage{subcaption} 
\usepackage{soul} 
\usepackage[hidelinks]{hyperref}
\usepackage[numbers,sort&compress]{natbib}
\usepackage[most]{tcolorbox}

\definecolor{lightblue}{HTML}{EEFBFF}
\definecolor{darkblue}{HTML}{C8EDF9}
\definecolor{darkerblue}{HTML}{2E4B87}
\definecolor{darkyellow}{HTML}{FEFAD4}
\definecolor{darkgrey}{HTML}{514E43}

\begin{document}

\title{LEVERAGING PREDICTION ENTROPY FOR AUTOMATIC PROMPT WEIGHTING IN ZERO-SHOT AUDIO-LANGUAGE CLASSIFICATION}
 
\name{\parbox{\textwidth}{\centering Karim El Khoury$^\dagger$$^{,1}$\qquad Maxime Zanella$^\dagger$$^{,1,2}$\qquad Tiffanie Godelaine$^\dagger$$^{,1}$\\
        \qquad \textit{Christophe De Vleeschouwer}$^{1}$\qquad \textit{Beno\^{i}t Macq}$^{1}$
\address{$^{1}$ICTEAM, UCLouvain, Belgium \qquad $^{2}$ILIA, UMons, Belgium}}}

\ninept
\maketitle
\renewcommand{\thefootnote}{\fnsymbol{footnote}}
\makeatletter
\renewcommand\@makefntext[1]{\noindent\normalfont#1} 
\footnotetext{$^\dagger$The authors have contributed equally to this work.
\vspace{1.5mm}\\Acknowledgments -- M.Z. is funded by the ARIAC (Walloon region grant No. 2010235). T.G. is funded by MedReSyst (EU-Wallonia 2021-2027 program).}
\makeatother

\begin{abstract}
Audio–language models have recently demonstrated strong zero-shot capabilities by leveraging natural-language supervision to classify audio events without labeled training data. Yet, their performance is highly sensitive to the wording of text prompts, with small variations leading to large fluctuations in accuracy. Prior work has mitigated this issue through prompt learning or prompt ensembling. However, these strategies either require annotated data or fail to account for the fact that some prompts may negatively impact performance. In this work, we present an entropy-guided prompt weighting approach that aims to find a robust combination of prompt contributions to maximize prediction confidence.  To this end, we formulate a tailored objective function that minimizes prediction entropy to yield new prompt weights, utilizing low-entropy as a proxy for high confidence. Our approach can be applied to individual samples or a batch of audio samples, requiring no additional labels and incurring negligible computational overhead. Experiments on five audio classification datasets covering environmental, urban, and vocal sounds, demonstrate consistent gains compared to classical prompt ensembling methods in a zero-shot setting, with accuracy improvements 5-times larger across the whole benchmark.


\end{abstract}

\begin{keywords}
Prompt Weighting,  Entropy Minimization, Zero-Shot Classification, Audio-Language Models
\end{keywords}

\section{Introduction}
\label{sec:intro}



Unlike approaches learning from predefined labels, Audio-Language Models (ALMs) utilize natural language as a supervision signal, which is more suitable for describing complex real-world audio recordings. This progress has been made possible by the development of self-supervised contrastive pre-training methods and the availability of large-scale text-audio datasets \cite{kim2019audiocaps, drossos2020clotho, wu2023large}. Contrastive pre-training on these large scale multimodal datasets is a powerful framework to obtain models that generalize well across diverse audio classification tasks. This design allows one to construct an audio classifier directly from textual descriptions of categories, without the need for additional training. Inspired by advancements in vision-language models such as CLIP \cite{radford2021learning}, CLAP \cite{clap22,clap23} is one of the first works to investigate its use in the audio field \cite{wu2023large,deshmukh2023pengi,yeh2023flap, xu2023blat}.\\


\begin{figure}[!t]
    \centering
    \includegraphics[width=\linewidth]{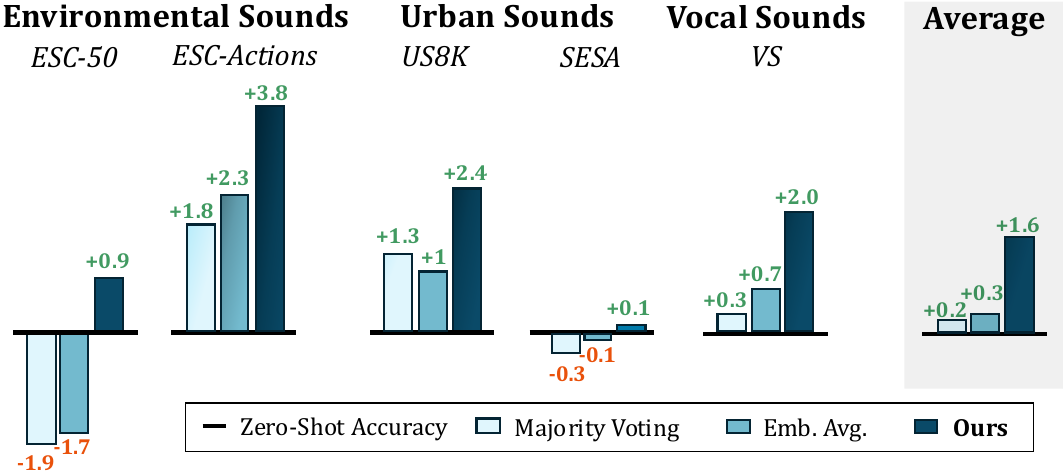}
    \caption{Summary of classification accuracy improvement over zero-shot prediction. Our approach, applied on the entirety of each dataset, shows consistent improvement over majority voting and embedding averaging ensemble methods.}
    \label{fig:summary_results}
\end{figure}

Despite ALMs having solid zero-shot classification accuracy \cite{clap22}, designing the right prompt for a given class becomes challenging, something even hindering final performance~\cite{hanif2024palm}. In fact, a slight change in a prompt can lead to vastly different predictions, potentially making the performance of a model unreliable. One solution to mitigate this problem is to rely on few-shot prompt learning \cite{hanif2024palm}. However, this approach can be computationally intensive and requires annotated samples. Another solution involves augmenting prompts using large language models \cite{11011255,olvera2024sound}.  While this is an interesting option, it is complementary to our line of work, as such methods can benefit from optimized prompt weighting to improve their overall performance. A third solution is to use an ensemble of templates \cite{clap22,clap23,deshmukh2023pengi,allingham2023simple} (Table \ref{tab:templates} shows common templates in the audio-language literature) to mitigate the sensitivity of models to the specific wording of a single prompt. Since selecting the right prompt among all templates is difficult, majority voting is often used on top of individual prompt predictions to make a final prediction. Another method is to simply use, for each class, the average of the embeddings of its individual prompts. These averaged embeddings are then used for classification.

While averaging text prompts is convenient, it is clear that some prompts perform significantly worse than others (see Figure \ref{fig:promptvariability}). In a purely zero-shot setting, it is however not straightforward to identify poorly performing prompts. 
Specifically, we hypothesize that appropriate weighting factors to average the text prompts should induce predictions with high confidence. Hence, we propose an entropy minimization objective, where the goal is to find a weighting vector, $\beta$, that aims to create a weighted combination of prompts that results in predictions with low entropy.

\textbf{Contributions}. We introduce a principled optimization framework that treats prompt ensembling as a problem of finding weights that maximize prediction confidence. This is achieved by minimizing the entropy of the final prediction in a purely zero-shot setting. Our approach allows to handle poorly performing prompts and overcomes the limitations of traditional ensembling methods, delivering consistent accuracy improvements. A summary of our results showing improvements over baseline ensemble methods on five audio classification benchmarks is shown in Figure \ref{fig:summary_results}.




\vspace{-1mm}

\section{Method}
\label{sec:method}






We present our iterative algorithm for finding prompt-weighting vector, $\beta$, by minimizing an objective function that drives for low-entropy (high-confidence) predictions. We regularize it against terms that prevent deviation from an initial zero-shot estimate and encourage smooth weight distribution.
\vspace{-1mm}
\subsection{Problem formulation}
We start by encoding each audio sample \( i \in \{1, \cdots, N_S\} \) with the ALM audio encoder to obtain a feature vector \( \mathbf{f}_i \in \mathbb{R}^{1 \times d} \). We get a set of text embeddings \( \mathbf{t}_{jk} \in \mathbb{R}^{1 \times d} \), where $\mathbf{t}_{jk}$ is obtained by combining template \( j \in \{1, \cdots, N_T\} \) with class \( k \in \{1, \cdots, N_C\} \) (e.g., \textit{This is a sound of $\{$class k$\}$}) and passing it into the ALM text encoder. We then compute the logit \( l_{ijk} \) of sample \( i \) for template \( j \) and class \( k \) as the cosine similarity\footnote{*Note that all vectors are normalized to the unit hypersphere, therefore the cosine similarity reduces to the dot product.}: 
\begin{equation}
l_{ijk} = \mathbf{f}_i \mathbf{t}_{jk}^T
\end{equation}


\noindent Next, we compute the average logit \( \bar{l}_{ik} \) of sample \( i \) for class \( k \) as a weighted average of the logits computed across all templates \( j \). This is given by:  
\begin{equation}
\bar{l}_{ik} = \sum_{j=1}^{N_T} \beta_j l_{ijk}
\end{equation}

where \( \beta \in \mathbb{R}^{1 \times N_T} \) is the weighting vector. In the uniform case, \( \beta = \frac{1}{N_T} \mathbf{1}_{N_T} \). By applying the softmax function, we obtain the prediction \( p_{ik} \) of sample \( i \) for class \( k \):  
\begin{equation}
p_{ik} = \frac{\exp(\bar{l}_{ik} / \tau)}{\sum_{l=1}^{N_C} \exp(\bar{l}_{il} / \tau)}
\label{eq:p}
\end{equation}

where \( \tau \) is a temperature scaling hyperparameter that controls the smoothness of the predictions.\\

In the case where $\beta$ is uniform, each template contributes equally to the final logit. However, we hypothesize that this method is sub-optimal, particularly when the classification accuracy across templates varies greatly, as observed in  Figure \ref{fig:promptvariability}. Hence, we aim to find the optimal contribution of each template, so that the model outputs confident predictions, i.e. low entropy $p_i$ vectors, across an entire batch of data samples.

\vspace{1.5mm}

\begin{tcolorbox}[width=\columnwidth, colframe=darkerblue, colback=white, boxsep=0.5mm, arc=2mm, left=2mm, right=2mm, top=2mm, bottom=2mm]
\centering
We formulate our optimization objective as finding\\ a weighting vector $\beta$ that produces confident\\ predictions on average.
\end{tcolorbox}



\newpage

\subsection{Objective variable}

Let the variables $\beta \in \mathbb{R}^{1 \times N_T}$ represent the weights for each of the $N_T$ distinct text prompt templates. This vector is constrained such that it lies in the probability simplex, enforcing the following conditions: $\sum_{j=1}^{N_T} \beta_{j} = 1$, and $\beta_{j} \ge 0 \quad \forall j$.

\vspace{-2mm}

\subsection{Objective function}

The goal is to minimize the objective function ${\cal L}(\beta)$ composed of three terms:  
\begin{align}
\hspace{-1.75mm} 
{\cal L}(\beta) &= 
\frac{1}{N_S} \sum_{i=1}^{N_S} 
\Big( 
\underbrace{H(p_i)}_{\shortstack{\scriptsize{\text{\textit{(i)}}} \\ \scriptsize{\text{\textit{Prediction}}} \\ \scriptsize{\text{\textit{confidence}}}}} 
+ \underbrace{\lambda_{zs} H(p_i, \hat{p}_i)}_{\shortstack{\scriptsize{\text{\textit{(ii)}}} \\ \scriptsize{\text{\textit{Zero-shot}}} \\ \scriptsize{\text{\textit{regularization
}}}}} 
\Big) 
- \underbrace{\lambda_{\beta} H(\beta)}_{\shortstack{\scriptsize{\text{\textit{(iii)}}} \\\scriptsize{\text{\textit{Entropy}}} \\ \scriptsize{\text{\textit{regularization}}}}} 
\label{eq:loss_function}
\end{align}

\begin{enumerate}[leftmargin=*]

\item[\textit{(i)}] \textit{Prediction confidence.} This term minimizes the entropy of the prediction vectors $p_i$ for each sample $i$, encouraging confident predictions across the $N_S$ samples.

\item[\textit{(ii)}] \textit{Zero-shot regularization.} The purpose of this term is to prevent the prediction vectors $p_i$ from deviating significantly from an initial prediction $\hat{p}_i$. Typically, we can obtain $\hat{p}_i$ with the single template \textit{This is a sound of $\{$class$\}$} which we name the zero-shot prediction.

\item[\textit{(iii)}] \textit{Entropy regularization.} This term is an entropy-based regularizer for the shared weight vector $\beta$, commonly used in the literature  \cite{martin2024transductive,zanella2024boosting,zanella2024boosting2,el2025enhancing}, which encourages smoother and non-sparse weight distributions. This term is outside the summation as $\beta$ is common to all samples. In addition, the convex penalty $-\lambda_{\beta} H(\beta)$ serves as an entropy barrier that implicitly enforces $\beta_j \ge 0$.

\end{enumerate}

\subsection{Optimization procedure}

The pseudo-code of the optimization procedure is outlined in Algorithm \ref{alg}. It should be noted that all embeddings $f$ and $t$ are computed only once before the optimization. At initialization, the prompt template weights are initialized with a uniform distribution and the zero-shot prediction vectors $\hat{p}$ are computed. The weights $\beta$ are updated according to an iterative fixed-point update rule derived from minimizing the objective function until convergence (i.e. solving $\frac{\partial \cal L}{\partial \beta_j}=0$): 
\begin{equation}
    \label{eq:b}
    \beta_j^{(t)} = \frac{\exp(R_j^{(t)}/\lambda_\beta)}{\sum_{l=1}^{N_T} \exp(R_l^{(t)}/\lambda_\beta)}
\end{equation}
with $R_j^{(t)}$ being interpreted as the sum of the contribution from each individual sample $i$:
\begin{equation}
\label{eq:R}
\begin{aligned}
R_j^{(t)} = \sum_{i=1}^{N_S} \Bigg[ 
& \sum_{k=1}^{N_C} p_{ik} \Big(\ln(p_{ik}) + H(p_i)\Big) l_{ijk}  \\
&\hspace{-1em} + \lambda_{zs} \sum_{k=1}^{N_C} p_{ik} \Big(\ln(\hat{p}_{ik}) - E_{p_i}[\ln(\hat{p}_{i})]\Big) l_{ijk} 
\Bigg]
\end{aligned}
\end{equation}

\vspace{2mm}

The contribution can be decomposed into two terms. The first term is linked to the entropy of $p_i$, while the second term evaluates the similarity with the zero-shot prediction, with $\lambda_{zs}$ balancing the contribution of the two terms. 

\begin{algorithm}[h]
\caption{Optimization Procedure}
\label{alg}
\SetKwComment{Comment}{$\triangleright$\ }{}
\KwIn{$\mathbf{f}$, $\mathbf{t}$, $\mathbf{\tau}$}
$\varepsilon \leftarrow 1\times10^{-6}$ \Comment*{\textcolor{darkerblue}{Stopping threshold}}
Initialize ${\beta^{(0)}} = \frac{1}{N_T} \mathbf{1}_{N_T}$ \Comment*{\textcolor{darkerblue}{Uniform weights}}
Compute $\hat{p}_i \quad \forall i$ \Comment*{\textcolor{darkerblue}{Zero-shot predictions}}
\While{$\lVert \beta^{(t)} - \beta^{(t-1)} \rVert_2 \geq \varepsilon$}{
    $t=t+1$\;
    Compute $p_i \quad \forall i$ \Comment*{\textcolor{darkerblue}{See Eq. \eqref{eq:p}}}
    Compute $R_j^{(t)} \quad \forall j$ \Comment*{\textcolor{darkerblue}{See Eq. \eqref{eq:R}}}
    Update ${\beta^{(t)}}$ \Comment*{\textcolor{darkerblue}{See Eq. \eqref{eq:b}}}
}
\textbf{return} ${\beta}$ \Comment*{\textcolor{darkerblue}{Optimized weights}}
\end{algorithm}

\section{Experimental Setup}

\subsection{Evaluation datasets}
We evaluate our method on five benchmark audio classification datasets spanning three different tasks: environmental, urban, and vocal sound classification. For environmental classification, we use ESC-50 \cite{esc50}, which contains $2k$ audio clips of 5 seconds each across 50 classes, and ESC-Actions \cite{esc50}, a subset of the previous dataset with 400 clips covering 10 action-related classes. For urban sound classification, we use US8K \cite{us8k}, comprising $\thicksim8k$ clips of $\thicksim4$ seconds each across 10 classes, and SESA \cite{sesa}, composed of $\thicksim600$ clips of $\thicksim30$ seconds each across 4 surveillance-related classes. For vocal sound classification, we employ VS \cite{vs}, consisting of $\thicksim21k$ clips grouped into 6 classes, with each clip being a 5-second recording. For all five datasets, we use the entire dataset as a single test set given that we are working in a zero-shot setting.
\vspace{-2mm}
\subsection{Text prompts}
We construct the set of templates using $N_T=35$ hand-crafted prompt templates commonly employed in the audio–language model literature \cite{clap22,clap23,deshmukh2023pengi,olvera2024sound}. These templates comprise a variety of sentence structures, detailed in Table~\ref{tab:templates}. The same set of templates is used across all five evaluation datasets. The relevance of this set of templates for prompt ensemble methods is highlighted by the observed high variability in accuracy across all five benchmark datasets (see Figure \ref{fig:promptvariability}).
\vspace{-2mm}
\subsection{Implementation details}
We use the original CLAP-2022~\cite{clap22} audio--language model to generate the audio embeddings $\mathbf{f}$ and all text embeddings $\mathbf{t}$. We set the base model logit scale to $33.3$, as per model specification. Concerning our method's hyperparameters, we fix $\lambda_{\beta}$ to 0.01 throughout our experiments. As for $\lambda_{zs}$, we set it to 100 for Experiment 1 (single-sample $\beta$) and 0.1 for Experiment 2 (dataset $\beta$). This choice is further discussed in Section \ref{sec:exp2}.
\vspace{-2mm}
\subsection{Baselines}
\label{sec:baselines}

We implement six different baseline prompt ensembling methods to compare our approach with:
\begin{enumerate}[leftmargin=*]
\vfill
\item[\textit{(i)}] \textit{Majority voting} consists of selecting the most frequently predicted class across the entire set of prompt templates for each sample. All prompt templates' votes are weighted equally.

\begin{table}[h]
\centering
\caption{Prompt templates used in our experiments. Class name should replace \{\}.}

\label{tab:templates}
\footnotesize 
\begin{tabular}{p{0.45\textwidth}}
\toprule
\textcolor{darkerblue}{\textbf{Declarative (This is...):}} This is a sound of \{\}, This is an audio of \{\}, This is an audio clip of \{\}, This is a sound clip of \{\}, This is an audio track of \{\}, This is a sound track of \{\}, This is an example of \{\}, This is \{\} \\
\midrule
\midrule
\textcolor{darkerblue}{\textbf{Descriptive (A/An...):}} A sound of \{\}, An audio of \{\}, A recording of \{\}, A sound recording of \{\}, An audio recording of \{\}, A sound clip of \{\}, An audio clip of \{\}, An audio track of \{\}, A sound track of \{\}, A sound snippet of \{\}, An audio snippet of \{\} \\
\midrule
\midrule
\textcolor{darkerblue}{\textbf{Imperative (Listen/Hear...):}} Listen to \{\}, Listen to the sound of \{\}, Listen to an audio of \{\}, Listen to a recording of \{\}, Listen to a sound recording of \{\}, Listen to an audio recording of \{\}, Hear the sound of \{\}, Hear an audio of \{\} \\
\midrule
\midrule

\textcolor{darkerblue}{\textbf{Direct Noun:}} Sound of \{\}, Audio of \{\}, Recording of \{\}, Sound recording of \{\}, Audio recording of \{\}, Audio clip of \{\} \\

\midrule
\midrule
\textcolor{darkerblue}{\textbf{Miscellaneous:}} I can hear \{\}, \{\} \\
\bottomrule
\end{tabular}
\end{table}

\vfill
\begin{figure}[h]
    \centering
    \includegraphics[width=\linewidth]{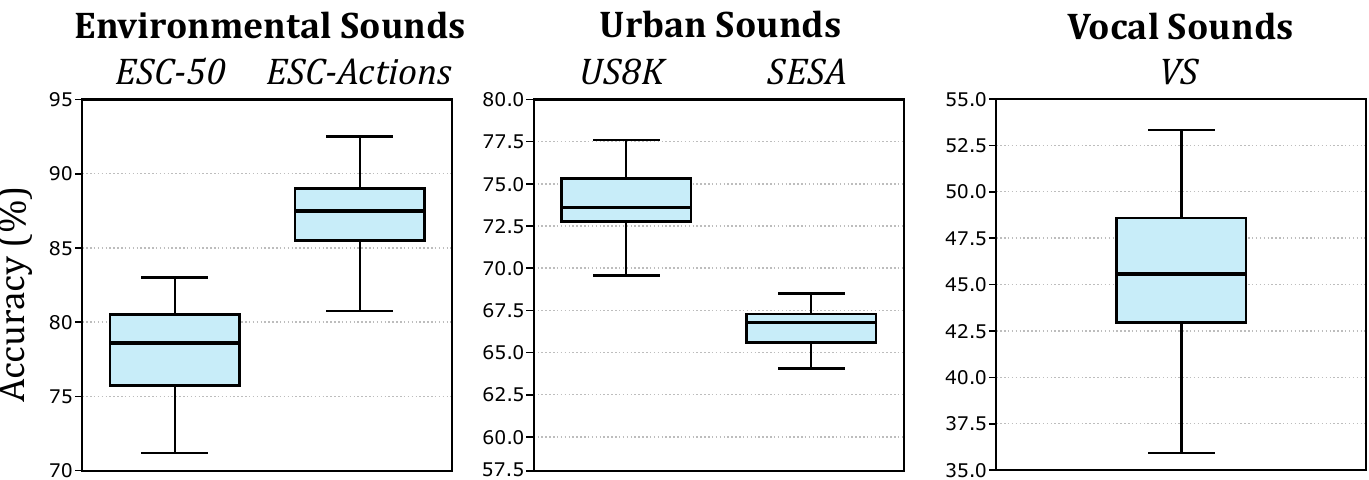}
    \caption{Distribution of accuracies obtained from 35 hand-crafted prompt templates across five audio datasets. It reveals substantial variability to prompt selection for all datasets.}

    \label{fig:promptvariability}
\end{figure}

\item[\textit{(ii)}] \textit{Majority voting \textit{with} entropy-weighting}, in contrast, weights each template's vote by the inverse of the entropy of its prediction vector. Thus, more confident template predictions have greater influence on the final classification.
\vspace{1.75mm}

\item[\textit{(iii)}] \textit{Majority voting \textit{with} pruning }consists of removing 50\% of the prompt templates with the highest entropy before performing a majority vote on the rest.
\vspace{1.75mm}

\item[\textit{(iv)}] \textit{Average text embedding} consists of calculating individual embeddings for each prompt template and then computing the normalized sum of these embeddings for each sample (i.e. ${\beta}$ is uniformly distributed). The resulting average text embedding is used for classification.
\vspace{1.75mm}

\begin{table*}[h]
\caption{Classification accuracy of our proposed approach on five audio classification datasets. We report three different settings: single-sample $\beta$ (i.e., $N_S=1$), dataset $\beta$ (i.e., $N_S$ equals the size of the dataset), and dataset $\beta$ with pruning (see Section \ref{sec:exp3} for more details) and compared against six baseline ensemble methods (see Section \ref{sec:baselines}).}
\vspace{-3mm}
\centering
\footnotesize
\label{tab:exp}
\begin{tabular}{p{6cm} !{\vrule width 1pt} c c  c c  c !{\vrule width 1pt} >{\centering\arraybackslash}c}
\multicolumn{1}{l!{\vrule width 1pt}}{\multirow{2}{*}{\textbf{Method}}} & 
\multicolumn{2}{c}{\textbf{Environmental Sounds}} & \multicolumn{2}{c}{\textbf{Urban Sounds}} & \multicolumn{1}{c!{\vrule width 1pt}}{\textbf{Vocal Sounds}}&
\multicolumn{1}{c}{\multirow{2}{*}{\textbf{Average}}} \\

\multicolumn{1}{c!{\vrule width 1pt}}{} &
\multicolumn{1}{c}{\textit{ESC-50}} &
\multicolumn{1}{c}{\textit{ESC-Actions}} &
\multicolumn{1}{c}{\textit{US8K}} &
\multicolumn{1}{c}{\textit{SESA}} &
\multicolumn{1}{c!{\vrule width 1pt}}{\textit{VS}} &
\multicolumn{1}{c}{} \\
\specialrule{.1em}{.0em}{.0em}

\hspace{-0.5em}{\scriptsize 
\color{darkerblue} \textbf{Zero-Shot Prediction}} & & & & & &  \\

\textit{``This is a sound of $\{$class$\}$"} & 82.6 & 87.7 & 75.0 & 66.7 & 46.9 & 71.8 \\
\rule{0pt}{2.5ex}
\hspace{-0.75em}{\scriptsize
\color{darkerblue} \textbf{Baseline Ensemble Methods}} & & & & & &  \\

\textit{Majority voting} & 80.7 & 89.5 & 76.3 & 66.4 & 47.2 & 72.0 \\
\textit{Majority voting \textit{with} entropy-weighting} & 81.2 & 90.3 & 75.1 & 66.6 & 47.4 & 72.1 \\
\textit{Majority voting \textit{with} pruning} & 82.9 & 90.3 & 75.1 & 67.1 & 46.0 & 72.3 \\

\rule{0pt}{2.5ex}\textit{Average text embedding} & 80.5 & 90.0 & 76.0 & 66.6 & 47.6 & 72.1 \\
\textit{Average text embedding} \textit{with} \textit{entropy-weighting}  & 80.6 & 90.0 & 76.2 & 66.6 & 47.6 & 72.2 \\
\textit{Average text embedding \textit{with} pruning}  & 82.7 & 90.3 & 75.1 & \underline{67.1} & 46.6 & 72.3 \\
\rule{0pt}{2.5ex}
\cellcolor{lightblue}\hspace{-0.75em}{\scriptsize 
\color{darkerblue} \textbf{Proposed Approach}} & \cellcolor{lightblue} & \cellcolor{lightblue} & \cellcolor{lightblue} & \cellcolor{lightblue} & \cellcolor{lightblue} & \cellcolor{lightblue} \\

\cellcolor{lightblue}\textit{Single-sample} \textit{$\beta$} & \cellcolor{lightblue}82.9 & \cellcolor{lightblue}90.0 & \cellcolor{lightblue}74.9 & \cellcolor{lightblue}\textbf{67.2} & \cellcolor{lightblue}47.7 & \cellcolor{lightblue}72.5 \\
\cellcolor{lightblue}\textit{Dataset} \textit{$\beta$} & \cellcolor{lightblue}\underline{83.5} & \cellcolor{lightblue}\underline{90.5} & \cellcolor{lightblue}\underline{77.3} & \cellcolor{lightblue}66.8 & \cellcolor{lightblue}\underline{47.9} & \cellcolor{lightblue}\underline{73.2} \\
\cellcolor{lightblue}\textit{Dataset} \textit{$\beta$} \textit{with} pruning & \cellcolor{lightblue}\textbf{83.5} & \cellcolor{lightblue}\textbf{91.5} & \cellcolor{lightblue}\textbf{77.4} & \cellcolor{lightblue}66.8 & \cellcolor{lightblue}\textbf{48.9} & \cellcolor{lightblue}\textbf{73.6} \\
\end{tabular}%

\end{table*}

\item[\textit{(v)}] \textit{Average text embedding \textit{with} entropy-weighting} modifies this process by weighting each text embedding in the normalized sum by the inverse of the entropy of its corresponding prediction vector. Therefore, text embeddings derived from confident templates contribute more significantly to the final embedding used for classification.
\vspace{1.75mm}
\item[\textit{(vi)}] \textit{Average text embedding \textit{with} pruning} consists of removing 50\% of the prompt templates with the highest entropy before calculating individual text embeddings for remaining prompt template and then computing the normalized sum of these embeddings for each sample. The resulting average text embedding is used for classification.
\end{enumerate}


\newpage

\section{Experiments}


\subsection{Experiment 1: Single-sample $\beta$}

In this experiment, we consider the scenario where we have access to one test sample at a time ($N_S = 1$) and, therefore, compute an individual weighting vector $\beta$ per sample. The results are shown in Table \ref{tab:exp}. We observe encouraging improvements, with an average gain of up to 0.7\% compared to the zero-shot setting, outperforming all baseline ensemble methods.
In particular, on ESC-50 we outperform the average text embedding baseline by $2.4\%$.
This highlights our capability to find an optimized $\beta$ weighting vector with just a \textit{single sample}, surpassing classical average weighting of text embeddings.
\vspace{-2mm}

\subsection{Experiment 2: Dataset $\beta$}
\label{sec:exp2}
In this experiment, we have access to the entire evaluation dataset and compute a common weighting vector, $\beta$, for all samples. We decrease the value of $\lambda_{zs}$ to 0.1, reducing the need for zero-shot regularization, as we compute $\beta$ with a larger number of samples. An ablation study on the $\lambda_{zs}$ parameter is shown in Table \ref{tab:param_optimization_two_rows} further motivating our choice. As seen in Table \ref{tab:exp}, we obtain average gains of $0.7\%$ compared to the single-sample $\beta$ optimization and average gains of $0.9\%$ over all baseline methods.

\vspace{-2mm}

\subsection{Experiment 3: Dataset $\beta$ with pruning}
\label{sec:exp3}
In this experiment, we perform multiple pruning cycles on the weighting vector $\beta$. During each cycle, we optimize $\beta$ before pruning it. The pruned $\beta$ vector is used to initialize the next cycle. We apply 4 cycles with a pruning percentage of 15\% at each cycle ending up with an overall pruning percentage of $\thicksim 50\%$. Additionally, we conduct an ablation study, depicted in Table \ref{tab:param_optimization}, to further validate our choice. As shown in Table \ref{tab:exp}, we achieve further average gains of 0.4\% compared to dataset-level $\beta$ optimization, resulting in cumulative average gains of 1.4\% across all baseline ensemble methods. Notably, the best performances are observed in 4 out of 5 datasets, with ESC-Actions achieving gains of up to 3.8\% over its zero-shot prediction. Note that even with multiple pruning cycles, the runtime of the proposed approach is comparable to the baseline methods, with a runtime of 0.2s (see Table \ref{tab:time}), which remains negligible compared to the time required to encode both audio and text features.

\newpage

\begin{table}[h]
\centering
\footnotesize

\caption{Average accuracy across all datasets for different values of $\lambda_{zs}$, highlighting the need for further regularization toward the zero-shot prediction in the single-sample optimization case. Choice highlighted in blue.}
\vspace{-2mm}
\label{tab:param_optimization_two_rows}
\setlength{\tabcolsep}{6pt}
\renewcommand{\arraystretch}{1}
\begin{tabular}{p{1.2cm}cccccc}
 & \multicolumn{6}{c}{\textbf{$\lambda_{zs}$}} \\
\cmidrule(lr){2-7} & \textit{0.01} & \textit{0.1} & \textit{1} & \textit{10} & \textit{100} & \textit{1000} \\
\specialrule{.1em}{.0em}{.0em}
\multicolumn{1}{c|}{Single-sample $\beta$} & 72.18 & 72.22 & 72.38 & 72.39 & \cellcolor{darkblue}\textbf{72.48} & 72.22 \\
\multicolumn{1}{c|}{Dataset $\beta$} & 72.79 & \cellcolor{darkblue}\textbf{73.19} & 71.33 & 70.84 & 70.62 & 70.61 \\
\end{tabular}
\end{table}

\vspace{-2mm}

\begin{table}[h]
\centering
\footnotesize
\caption{Average accuracy across all datasets for different pruning percentages per cycle and total cycles. Choice highlighted in blue.}
\vspace{-2mm}

\label{tab:param_optimization}
\setlength{\tabcolsep}{6pt}
\renewcommand{\arraystretch}{1}
\begin{tabular}{p{2.2cm}ccccc}

 & \multicolumn{5}{c}{\textbf{Total Cycles}} \\
\cmidrule(lr){2-6}
\textbf{Pruning\% / Cycle} & \textit{1} & \textit{2} & \textit{3} & \cellcolor{lightblue}\textit{4} & \textit{5} \\
\specialrule{.1em}{.0em}{.0em}
\multicolumn{1}{c|}{\textit{5}}  & 73.26 & 72.92 & 73.13 & \cellcolor{lightblue}73.33 & 73.30 \\
\multicolumn{1}{c|}{\textit{10}} & 72.96 & 73.31 & 73.54 & \cellcolor{lightblue}73.42 & 73.45 \\
\multicolumn{1}{>{\centering\arraybackslash}p{2.2cm}|}{\cellcolor{lightblue}\textit{15}}
                        & \cellcolor{lightblue}73.22
                        & \cellcolor{lightblue}73.35
                        & \cellcolor{lightblue}73.52
                        & \cellcolor{darkblue}\textbf{73.60}
                        & \cellcolor{lightblue}73.56 \\
\multicolumn{1}{c|}{\textit{20}} & 73.37 & 73.45 & 73.43 & \cellcolor{lightblue}73.34 & 73.20 \\
\multicolumn{1}{c|}{\textit{25}} & 73.46 & 73.42 & 73.38 & \cellcolor{lightblue}{72.21} & 72.12
\end{tabular}
\end{table}

\vspace{-2mm}
\begin{table}[h]
    \centering
    \footnotesize

    \caption{Runtime of our approach compared to feature extraction and baseline methods. Evaluation was performed with 35 templates on a 24 GB NVIDIA GeForce RTX 4090 GPU using the VS dataset ($\thicksim 21k$ audio samples split into six classes). }
\vspace{-2mm}

    \label{tab:time}
    \begin{tabular}{lccc}
         \textbf{} & Features encoding & Baselines &  Proposed Approach\\
        \specialrule{.1em}{.0em}{.0em}
         \multicolumn{1}{l}{Runtime} &\multicolumn{1}{|c}{$\thicksim 2 \text{ min}$} & $\thicksim 0 \text{ s}$ &   \cellcolor{darkblue}$\thicksim 0.2 \text{ s}$
    \end{tabular}
\end{table}


\section{Conclusion}
\label{sec:conclu}

We propose an entropy-guided prompt weighting algorithm for zero-shot audio–language classification that strategically combines embeddings from multiple templates, using low-entropy predictions as a proxy for high confidence. We demonstrate that this approach consistently outperforms baseline ensemble methods across five benchmark audio classification datasets. The proposed method offers a practical solution that can be integrated into existing audio-language model zero-shot classification pipelines, alleviating the need for manual prompt engineering when encountering new datasets.


 

\newpage

\bibliographystyle{IEEEbib}
\bibliography{strings,refs}

\begin{thebibliography}{10}

\bibitem{kim2019audiocaps}
Chris~Dongjoo Kim, Byeongchang Kim, Hyunmin Lee, and Gunhee Kim,
\newblock ``Audiocaps: Generating captions for audios in the wild,''
\newblock in {\em Proceedings of the 2019 Conference of the North American Chapter of the Association for Computational Linguistics: Human Language Technologies, Volume 1 (Long and Short Papers)}, 2019, pp. 119--132.

\bibitem{drossos2020clotho}
Konstantinos Drossos, Samuel Lipping, and Tuomas Virtanen,
\newblock ``Clotho: An audio captioning dataset,''
\newblock in {\em ICASSP 2020-2020 IEEE International Conference on Acoustics, Speech and Signal Processing (ICASSP)}. IEEE, 2020, pp. 736--740.

\bibitem{wu2023large}
Yusong Wu, Ke~Chen, Tianyu Zhang, Yuchen Hui, Taylor Berg-Kirkpatrick, and Shlomo Dubnov,
\newblock ``Large-scale contrastive language-audio pretraining with feature fusion and keyword-to-caption augmentation,''
\newblock in {\em ICASSP 2023-2023 IEEE International Conference on Acoustics, Speech and Signal Processing (ICASSP)}. IEEE, 2023, pp. 1--5.

\bibitem{radford2021learning}
Alec Radford, Jong~Wook Kim, Chris Hallacy, Aditya Ramesh, Gabriel Goh, Sandhini Agarwal, Girish Sastry, Amanda Askell, Pamela Mishkin, Jack Clark, et~al.,
\newblock ``Learning transferable visual models from natural language supervision,''
\newblock in {\em International conference on machine learning}. ICML, 2021, pp. 8748--8763.

\bibitem{clap22}
Benjamin Elizalde, Soham Deshmukh, Mahmoud~Al Ismail, and Huaming Wang,
\newblock ``Clap learning audio concepts from natural language supervision,''
\newblock in {\em ICASSP 2023 - 2023 IEEE International Conference on Acoustics, Speech and Signal Processing (ICASSP)}, 2023, pp. 1--5.

\bibitem{clap23}
Benjamin Elizalde, Soham Deshmukh, and Huaming Wang,
\newblock ``Natural language supervision for general-purpose audio representations,''
\newblock in {\em ICASSP 2024-2024 IEEE International Conference on Acoustics, Speech and Signal Processing (ICASSP)}. IEEE, 2024, pp. 336--340.

\bibitem{deshmukh2023pengi}
Soham Deshmukh, Benjamin Elizalde, Rita Singh, and Huaming Wang,
\newblock ``Pengi: An audio language model for audio tasks,''
\newblock {\em Advances in Neural Information Processing Systems}, vol. 36, pp. 18090--18108, 2023.

\bibitem{yeh2023flap}
Ching-Feng Yeh, Po-Yao Huang, Vasu Sharma, Shang-Wen Li, and Gargi Gosh,
\newblock ``Flap: Fast language-audio pre-training,''
\newblock in {\em 2023 IEEE Automatic Speech Recognition and Understanding Workshop (ASRU)}. IEEE, 2023, pp. 1--8.

\bibitem{xu2023blat}
Xuenan Xu, Zhiling Zhang, Zelin Zhou, Pingyue Zhang, Zeyu Xie, Mengyue Wu, and Kenny~Q Zhu,
\newblock ``Blat: Bootstrapping language-audio pre-training based on audioset tag-guided synthetic data,''
\newblock in {\em Proceedings of the 31st ACM International Conference on Multimedia}, 2023, pp. 2756--2764.

\bibitem{hanif2024palm}
Asif Hanif, Maha~Tufail Agro, Mohammad~Areeb Qazi, and Hanan Aldarmaki,
\newblock ``Palm: Few-shot prompt learning for audio language models,''
\newblock {\em arXiv preprint arXiv:2409.19806}, 2024.

\bibitem{11011255}
Nishit Anand, Ashish Seth, Ramani Duraiswami, and Dinesh Manocha,
\newblock ``Tspe: Task-specific prompt ensemble for improved zero-shot audio classification,''
\newblock in {\em 2025 IEEE International Conference on Acoustics, Speech, and Signal Processing Workshops (ICASSPW)}, 2025, pp. 1--5.

\bibitem{olvera2024sound}
Michel Olvera, Paraskevas Stamatiadis, and Slim Essid,
\newblock ``A sound description: Exploring prompt templates and class descriptions to enhance zero-shot audio classification,''
\newblock {\em arXiv preprint arXiv:2409.13676}, 2024.

\bibitem{allingham2023simple}
James~Urquhart Allingham, Jie Ren, Michael~W Dusenberry, Xiuye Gu, Yin Cui, Dustin Tran, Jeremiah~Zhe Liu, and Balaji Lakshminarayanan,
\newblock ``A simple zero-shot prompt weighting technique to improve prompt ensembling in text-image models,''
\newblock in {\em International Conference on Machine Learning}. ICML, 2023, pp. 547--568.

\bibitem{martin2024transductive}
S{\'e}gol{\`e}ne Martin, Yunshi Huang, Fereshteh Shakeri, Jean-Christophe Pesquet, and Ismail Ben~Ayed,
\newblock ``Transductive zero-shot and few-shot clip,''
\newblock in {\em Proceedings of the IEEE/CVF Conference on Computer Vision and Pattern Recognition}, 2024, pp. 28816--28826.

\bibitem{zanella2024boosting}
Maxime Zanella, Beno{\^\i}t G{\'e}rin, and Ismail Ayed,
\newblock ``Boosting vision-language models with transduction,''
\newblock {\em Advances in Neural Information Processing Systems}, vol. 37, pp. 62223--62256, 2024.

\bibitem{zanella2024boosting2}
Maxime Zanella, Fereshteh Shakeri, Yunshi Huang, Houda Bahig, and Ismail~Ben Ayed,
\newblock ``Boosting vision-language models for histopathology classification: Predict all at once,''
\newblock in {\em International Workshop on Foundation Models for General Medical AI}. Springer, 2024, pp. 153--162.

\bibitem{el2025enhancing}
Karim El~Khoury, Maxime Zanella, Beno{\^\i}t G{\'e}rin, Tiffanie Godelaine, Beno{\^\i}t Macq, Sa{\"\i}d Mahmoudi, Christophe De~Vleeschouwer, and Ismail~Ben Ayed,
\newblock ``Enhancing remote sensing vision-language models for zero-shot scene classification,''
\newblock in {\em ICASSP 2025-2025 IEEE International Conference on Acoustics, Speech and Signal Processing (ICASSP)}. IEEE, 2025, pp. 1--5.

\bibitem{esc50}
Karol~J. Piczak,
\newblock ``{ESC}: {Dataset} for {Environmental Sound Classification},''
\newblock in {\em Proceedings of the 23rd {Annual ACM Conference} on {Multimedia}}. pp. 1015--1018, {ACM Press}.

\bibitem{us8k}
Justin Salamon, Christopher Jacoby, and Juan~Pablo Bello,
\newblock ``A dataset and taxonomy for urban sound research,''
\newblock in {\em Proceedings of the 22nd ACM international conference on Multimedia}, 2014, pp. 1041--1044.

\bibitem{sesa}
Tito Spadini,
\newblock ``Sound events for surveillance applications,'' oct 2019,
\newblock October 2019.

\bibitem{vs}
Yuan Gong, Jin Yu, and James Glass,
\newblock ``Vocalsound: A dataset for improving human vocal sounds recognition,''
\newblock in {\em ICASSP 2022-2022 IEEE International Conference on Acoustics, Speech and Signal Processing (ICASSP)}. IEEE, 2022, pp. 151--155.

\end{thebibliography}

\end{document}